\documentclass[
superscriptaddress,
preprint,
nobibnotes,
aip,
apl,
floatfix,
]{revtex4-2}
\usepackage{amsfonts,amsmath,dsfont}
\usepackage{amssymb, bm, mathrsfs}
\usepackage{graphicx,xcolor,color,soul,natbib,url}
\usepackage[labelformat=simple]{subcaption}
\usepackage{physics,chemformula}
\usepackage[colorlinks=true,citecolor=blue]{hyperref}
\usepackage[capitalise]{cleveref}
\usepackage{appendix, comment}
\usepackage[mathlines]{lineno}
\usepackage{ragged2e,cancel,relsize}
 
\DeclareCaptionJustification{justified}{\justifying}
\newcommand{\lr}[1]{\left({#1}\right)}
\newcommand{\mbf}[1]{\mathbf{#1}}
\begin{document}
\title{Quantum Transport and Shot Noise in Two-Dimensional Semi-Dirac System}

\author{Wei Jie Chan}
\affiliation{Science, Mathematics and Technology (SMT), Singapore University of Technology and Design (SUTD), 8 Somapah Road, Singapore 487372}

\author{L. K. Ang}
\thanks{Authors to whom correspondence should be addressed: ricky\_ang@sutd.edu.sg and yeesin\_ang@sutd.edu.sg}
\affiliation{Science, Mathematics and Technology (SMT), Singapore University of Technology and Design (SUTD), 8 Somapah Road, Singapore 487372}

\author{Yee Sin Ang}
\thanks{Authors to whom correspondence should be addressed: ricky\_ang@sutd.edu.sg and yeesin\_ang@sutd.edu.sg}
\affiliation{Science, Mathematics and Technology (SMT), Singapore University of Technology and Design (SUTD), 8 Somapah Road, Singapore 487372}

\begin{abstract}
Two-dimensional ($2$D) semi-Dirac systems, such as $2$D black phosphorus and arsenene, can exhibit a rich topological phase transition between insulating, semi-Dirac, and band inversion phases when subjected to an external modulation.
How these phase transitions manifest within the quantum transport and shot noise signatures remain an open question thus far.
Here, we show that the Fano factor converges to the universal $F\approx0.179$ at the semi-Dirac phase, and transits between the sub-Poissonian ($F\approx1/3$) and the Poissonian shot noise ($F\approx1$) limit at the band inversion and the insulating phase, respectively.
Furthermore, the conductance of $2$D semi-Dirac system converges to the contrasting limit of $G/G_0 \rightarrow 1/d$ and $G/G_0 \rightarrow0$ at the band inversion and the insulating phases, respectively.
The quantum tunneling spectra exhibits a peculiar coexistence of massless and massive Dirac quasiparticles in the band inversion regime, thus providing a versatile sandbox to study the tunneling behavior of various Dirac quasiparticles.
These findings reveal the rich interplay between band topology and quantum transport signatures, which may serve as smoking gun signatures for the experimental studies of semi-Dirac systems near topological phase transition. 
\end{abstract}
\maketitle

Shot noise is generated by electrical current fluctuations arising from the discrete nature of charge carriers. Shot noise can be used as an indicator of the correlation between charged carriers \cite{levitov1993charge,Blanter2000,Ghosh2022,Chevallier2010}.
A well-known characterization of shot noise is the Fano factor, $F$, which is defined as the ratio between the actual shot noise and the Poisson shot noise with $F=1$. 
Prominent examples of systems with $F\neq 1$ include the super-Poissonian shot noise with $F>1$ in zero-dimensional quantum dots \cite{Onac2006,Harabula2018} and the sub-Poissonian shot noise with $F=1/3$ in both disordered conductors \cite{Nagaev1992,Beenakker1992} and graphene \cite{Tworzydo2006,Danneau2009,Danneau2008}.
Furthermore, the well-celebrated maximal Fano factor value in graphene is shown to be associated with the minimal conductivity in orders of $e^2/\hbar$ \cite{Tworzydo2006}.
Importantly, shot noise provides a useful tool to probe the quantum transport properties of an electronic system \cite{Beenakker2003}, and has been widely employed in the experimental studies of graphene and their heterostructures \cite{Sahu2019,kumada2015shot}.

Two-dimensional ($2$D) semi-Dirac material (SDM) represents an interesting system that simultaneously host linear (relativistic) energy dispersion in one direction and parabolic (nonrelativistic) energy dispersion in the orthogonal direction \cite{Banerjee2009,Banerjee2012,Huang2015}.
SDMs has been realized in a large variety of systems, including \ch{(TiO2)_m}/\ch{(VO2)_n} nanostructure \cite{Pardo2009}, strained organic salt \cite{Katayama2006}, photonic crystals \cite{Wu2014}, \ch{Bi_{1-x}Sb_x} \cite{Tang2012}, striped boron sheet \cite{Zhang2017}, on surface states of topological insulators \cite{Li2011,Zhai2011}, or in non-centrosymmetric systems \cite{Park2017} such phosphorus-based materials \cite{Baik2015,Ghosh2016,Liu2015,Kim2017,Adhikary2021,WangJAP2015,Yarmohammadi2020}, monolayer arsenene \cite{Wang2016}, silicene oxide \cite{Zhong2017}, and polariton lattice \cite{Real2020} and also shown in $\alpha$-dice lattice \cite{Carbotte2019,Mandhour2020,Illes2017} with higher pseudospin.
The electronic transport and shot noise of SDM have been studied extensively in previous works, which establish a Fano factor of $F=1$ and $F\approx 0.179$ at the band insulator phase with nonzero band gap and at the semimetallic gapless limit, respectively  \cite{Ghasemian2021,Zhai2014,Illes2017,Betancur2019,Choi2021,Li2017,Jung2020,Ang2017,Saha2017,Rostamzadeh2022}.  
Nevertheless, 2D SDM can undergo complex topological phase transitions. Beyond the band insulating and the semimetallic regime, SDM can exhibit a band inversion phase \cite{Park2017} in which two distinct Dirac cones emerge, thus rendering the band-inverted SDM a strong potential for valleytronic device applications \cite{Ang2017}. 
Nevertheless, the shot noise and conductance signatures of band-inverted SDM remains an open question thus far.

\begin{figure}
    \centering
    \includegraphics[width=0.45\textwidth]{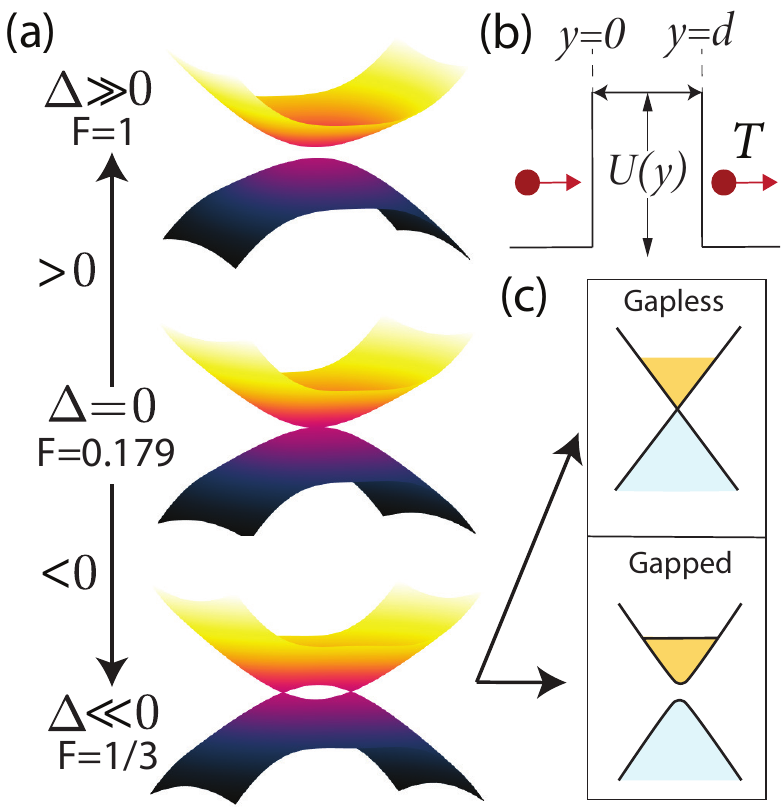}
    \caption{Illustration of the insulating ($\Delta>0$), semi-Dirac ($\Delta=0$) and band inversion ($\Delta<0$) phase tuned by $\Delta$ in \cref{eqn:H_full} with its corresponding Fano factors in (a). 
    Illustration of chiral tunneling across a potential barrier with width $d$ and transmission probability $T$ along the linear dispersion ($k_y$) in (b). 
    The band inversion phase in (a) can co-exhibit both gapless/gapped band crossings along $k_y$ depending on its transverse momenta, $k_x$ in (c).}
    \label{fig:1}
\end{figure}

In this work, we study the quantum transport of SDM near the topological phase transitions. 
Focusing on the quantum transport occuring along the relativistic direction - not covered in previous quantum transport study \cite{Ang2017}, we observe the intriguing coexistence of massless \cite{Tworzydo2006,Katsnelson2006} and massive Dirac fermions in the tunneling spectra at fixed transport channel at all quasiparticle energies, which is distinctive from that of bilayer graphene \cite{He2013} in which the massive and massless Dirac quasiparticle occurs at various transversal momenta at different quasiparticle energies. 
We further calculate the conductance and Fano factors of $2$D SDMs as the band topology changes continuously from band insulating to band inversion phases. 
Remarkably, we found that the Fano factor converges to sub-Poissonian shot noise with $F\approx 1/3$ and to Poissonian shot noise with $F\approx 1$ in the band inversion and insulating phase, respectively [\cref{fig:1}(a)]. 
Such shot noise signatures have negligible thermal noise contributions, even at room temperature.
Our findings reveal the exotic quantum transport behavior and the shot noise signatures of 2D SDM at various phases, thus uncovering shot noise as a useful tool in probing the band topology of $2$D SDM.

$2$D SDM can be described by a two-band effective Hamiltonian \cite{Montambaux2009,Montambaux2009_2}, 
\begin{align}
    \hat{H} = \lr{\alpha k^2_x + \Delta}\sigma_x + \beta k_y \sigma_y, \label{eqn:H_full}    
\end{align}
with $\alpha = \hbar^2/(2m^*)$ and $\beta = \hbar v_y$, where $m^*$ and $v_y$ being the effective mass along $\hat{x}$ and Fermi velocity along $\hat{y}$, respectively.
A phase transition parameter, $\Delta$, acts as a perturbation factor that continuously changes the band topology from band insulating phase ($\Delta >0$), semi-Dirac phase ($\Delta = 0$) to band inversion phase ($\Delta <0$).
We can nondimensionlise \cref{eqn:H_full} by defining the characteristic momentum and energy ($\hbar k_0 = 2m^*v_y$ and $\varepsilon_0 = \hbar k_0v_y$) to
obtain $\hat{\mathcal{H}} = \lr{k^2_x+\Delta}\sigma_x + k_y\sigma_y$ which has the following energy dispersion
\begin{align}
    \varepsilon_{\mbf{k}} = \pm\sqrt{\lr{k^2_x+\Delta}^2+k^2_y},
    \label{eqn:Gen_E}
\end{align}
where $\pm$ indicates conduction/valence ($+$/$-$) bands.
This parameter has been utilized for directional dependent transport \cite{Mawrie2019,Zhai2014,Zhou2021,Zhang2022,Kim2015,Cunha2022} and phase-dependent transport \cite{Sriluckshmy2018,Carbotte2019,Saha2016,Jung2020}.
Owing to the presence of two inequivalent valleys in the band inversion phase ($\Delta<0$), band-inverted $2$D SDMs have been widely studied for potential applications in valleytronics \cite{Jung2020,Ang2017,Rostamzadeh2022,Saha2017}.

\begin{figure}
    \centering
    \includegraphics[width=0.45\textwidth]{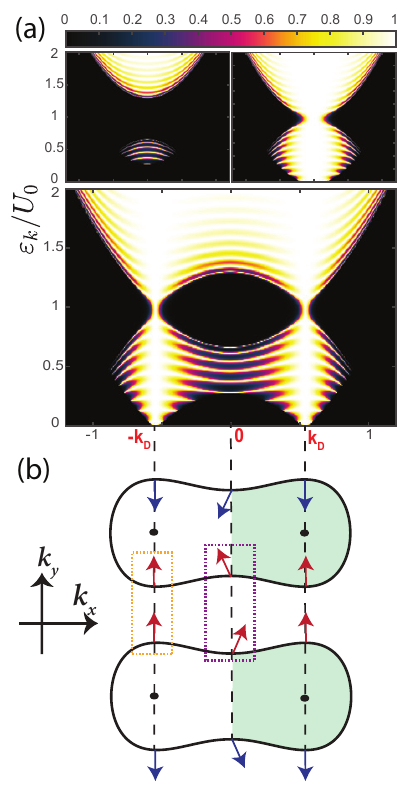}
    \caption{
    Transmission probability $T$ for $\Delta = 0.3$, $0$ and $-0.3$ are shown in the top left/top right/bottom panels in (a).
    Pseudospin vectors from \cref{eqn:S} are shown in (b) with red/blue arrows indicating the the direction of $S_y$. 
    }
    \label{fig:2}
\end{figure}

From \cref{fig:1}(a), the SDM ($\Delta = 0$) with a semi-Dirac point at $(k_x,k_y)=(0,0)$ can become a band insulator ($\Delta > 0$) or exhibit a band inversion ($\Delta < 0$) phase with band crossings at $(k_x,k_y) =(\pm k_D,0)$, with $k_D =\sqrt{\abs{\Delta}}$ \cite{Banerjee2009,Huang2015,Park2017}.
Interestingly, \cref{eqn:H_full} in the band inversion regime can be decomposed into either the $1$D massless and massive Dirac Hamiltonian [\cref{fig:1}(c)] along the $k_y$ direction at $k_x = \pm k_D$ and at $k_x=0$, respectively,
\begin{align}
    \hat{H}(\Delta<0)= 
    \begin{cases}
        k_y \sigma_y,&k_x = \pm k_D;\\
       \Delta\sigma_x + k_y \sigma_y,& k_x =0.
    \end{cases}\label{eqn:H_BandInv}
\end{align}
The quantum transport along the $k_y$ direction is thus expected to exhibit a mixture of massive and massive Dirac quasiparticles dictated by \cref{eqn:H_BandInv}. 

The transmission probability $T$ can be obtained by considering a scattering potential along $k_y$ [\cref{fig:1}(b)] with $U(y) = \mathcal{U}_0(\Theta(y) - \Theta(y-d))$, where $U_0 \equiv \mathcal{U}_0/\varepsilon_0$ and $d \equiv d_0 k_0$ are the dimensionless potential height/barrier width respectively.
In the $L_x \gg d$ limit, the Hamiltonian decouples into a $1$D eigenvalue equation along $k_y \rightarrow -i\partial_y$, with 
\begin{align}
    \Psi_{j} = A_{j}
    \begin{pmatrix}
    \varepsilon_{j}\\k^2_x + \Delta + ik_{j}
    \end{pmatrix}e^{ik_jy} + B_j 
    \begin{pmatrix}
    \varepsilon_{j}\\k^2_x + \Delta - ik_{j}
    \end{pmatrix}e^{-ik_jy},
    \label{eqn:wavefun}
\end{align}
where the index $j$ denotes the L (left), B (barrier), and (R) right region.
The energy and wavevector with index $j$ denotes $\varepsilon_{L/R} = \varepsilon_\mbf{k}$, $\varepsilon_B = \varepsilon_\mbf{k}-U_0 \equiv \varepsilon_\mbf{q}$ with $k_{L/R} =k_y= \lambda\sqrt{\varepsilon^2_\mbf{k}-(k^2_x+\Delta)^2}$, $k_B =q_y =\lambda'\sqrt{\varepsilon^2_\mbf{q}-(k^2_x+\Delta)^2}$, $\lambda=\text{sgn}(\varepsilon_\mbf{k})$ and $\lambda'=\text{sgn}(\varepsilon_\mbf{q})$.
For the left incident wavefunction, $A_L = 1$ and $B_R = 0$ are enforced with the transmission coefficient $t = A_R$.
For only forward-moving electronic states in region R, the wavevectors are enforced in \cref{eqn:wavefun} through $k_y>0$ and $q_y<0$.
For \textit{n-p-n} junction, the conservation of current (non-dimensionlized), $\hat{J}_i + \hat{J}_r = \hat{J}_t$ with $\hat{J}_y =\frac{\hbar}{2} \Psi^\dagger \hat{\sigma}_y\Psi =\frac{\hbar}{2} \Psi^\dagger \hat{\sigma}_y\Psi,$ gives us the relation $\abs{r}^2 + \abs{t}^2 = 1$.
By matching the boundary conditions at $y = 0$ and $d$, we obtain $T=\abs{t}^2$ as
\begin{widetext}
\begin{align}
    T=\frac{4\varepsilon^2_\mbf{k} \varepsilon^2_\mbf{q} k^2_y q^2_y}{4\varepsilon^2_\mbf{k} \varepsilon_\mbf{q}^2 k^2_y q^2_y \cos^2(\lambda' q_y d)+ \sin^2(\lambda' q_yd) [(k^2_x +\Delta)^2U^2_0+\varepsilon^2_\mbf{k}q^2_y + \varepsilon^2_\mbf{q}k^2_y]^2}, 
    \label{eqn:Trans_NPN}
\end{align}
\end{widetext} 
in agreement with a band insulator for $\Delta>0$ [top left panel of \cref{fig:2}(a)] and a SDM \cite{Zhai2014} for $\Delta =0$ by noticing that $\tan\phi=\lr{k^2_x+\Delta}/k_y$ and $\tan\theta=\lr{k^2_x+\Delta}/q_y$ [top right panel of \cref{fig:2}(a)].
Remarkably, the rotational invariant of \cref{eqn:Gen_E} around $\hat{z}$, i.e. $[H,R_z]=0$ with $R_z$ being a rotation operator about the $\hat{z}$, allows \cref{eqn:Trans_NPN} to fully capture the Klein tunneling behavior, as evident under rotation and incident angle \cite{Ghasemian2021,Li2017}.
However, this cannot be generalized along the parabolic direction due to intervalley scattering \cite{Zhai2014,Ang2017,Ghasemian2021,Betancur2019}.

We now focus on the band inversion phase for $\Delta < 0$ [bottom panel of \cref{fig:2}(a)].
The tunneling behavior \cite{Mandhour2020} in \cref{eqn:Trans_NPN} at different transverse momenta, $k_x$, can be understood from the pseudospin texture \cite{Ang2017} (non-dimensionalized) shown in \cref{fig:2}(b) and is given by
\begin{align}
    \mbf{S} = \Psi^\dagger \bm{\sigma} \Psi = \lr{\frac{k^2_x + \Delta}{\varepsilon_\mbf{k}},\lambda\frac{k_y}{\varepsilon_\mbf{k}}}.
    \label{eqn:S}
\end{align} 
As $T$ is symmetric about $k_x$ [green shading in the lower half contours], it suffices to analyze the non-positive transverse momenta with $k_x \leq 0$.
At a Dirac point $k_x = -k_D$, the conservation of pseudospin across the potential barrier [orange dotted box] enables Klein tunneling at normal incidence with $T=1$ [\cref{fig:2}(a)], which resembles the $1$D tunneling of massless Dirac fermions at $k_x = -k_D$ in \cref{eqn:H_full}.
Similarly, transmission resonances occur at $0\leq k_x < k_D$ [purple box in \cref{fig:2}(a)], which resembles the $1$D tunneling of gapped massive Dirac fermions.

As illustrated in \cref{eqn:H_BandInv}, there exist two special transport channel at $k_x = k_D$ in which the quasiparticles tunnel with full transmission. 
Such Klein tunneling behavior arises because of the $1$D gapless massive Dirac Hamiltonian at $k_x = k_D$ [see \cref{eqn:H_BandInv}].
At $k_x = 0$, the effective $1$D Hamiltonian [\cref{eqn:H_BandInv}] reduces to that of a gapped Dirac quasiparticle. 
The tunneling spectra at the transport channel of $k_x=0$ thus deviates from perfect Klein tunneling and exhibits oscillations instead.
It should be noted that the coexistence of massive and massless Dirac fermions have also been reported in twisted bilayer graphene \cite{He2013}. 
However, in twisted bilayer graphene, the massive and massless Dirac quasiparticles occur at different transport channel (i.e. transversal momenta) dependent on the energy of the quasiparticles. 
This aspect is in stark contrast to the case of $2$D SDM studied here, in which the massless and massive Dirac quasiparticles occur at \emph{fixed} transport channels of $k_x = k_D$ and $k_x = 0$, respectively. 

\begin{figure} [t]
    \centering
    \includegraphics[width=0.485\textwidth]{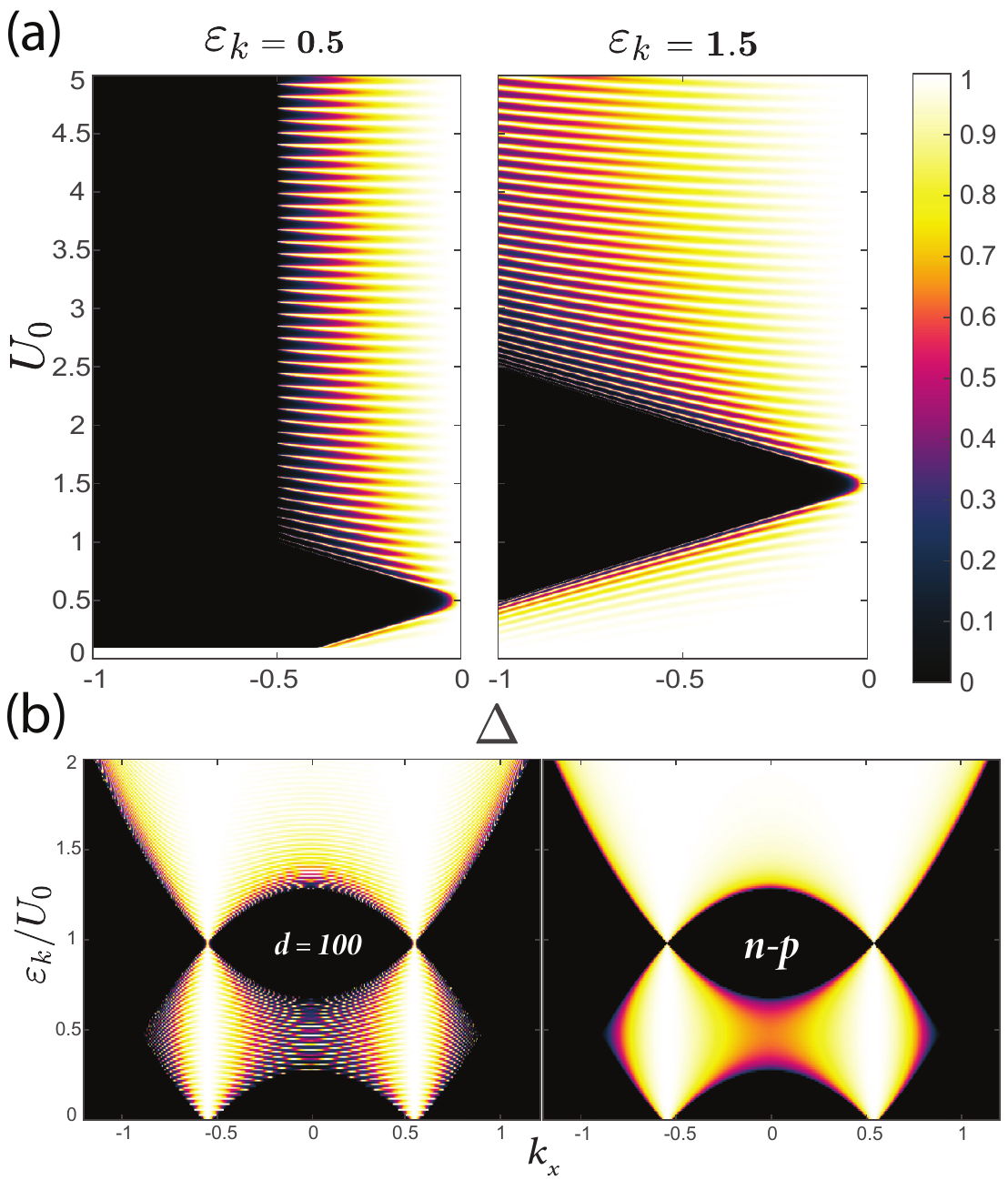}
    \caption{
    The relationship of $U_0$ against $\Delta$ on $T$ are plotted in (a) with both $\varepsilon_\mbf{k}=0.5$ and $1.5$.
    The effects of potential barrier width $d$ on $T$ are shown in (b) for $d=100$ and in the \textit{n-p} limit at $\Delta = -0.3$.}
    \label{fig:3}
\end{figure}

The modulation of pseudogap through $U_0$ and $\Delta$ is shown in \cref{fig:3}(a).
While the full transmission, oscillatory and pseudogap regions are expected from Klein tunneling \cite{Tworzydo2006,He2013}, an abrupt cutoff for $\varepsilon_\mbf{k} = 0.5$ at $\Delta =0.5$ is observed. 
This is attributed by the crossover of the incident wavefunction from propagating to evanescent when $k_y = \sqrt{\varepsilon^2_\mbf{k} - \Delta^2}$ becomes imaginary at $k_x = 0$.
This implies that the cutoff for $\varepsilon_\mbf{k} = 1.5$ appears only at $\Delta = -1.5$. 
Hence, the decoupling of the band inversion phase with $\Delta \ll 0$, in the semi-Dirac literatures \cite{Montambaux2009,Montambaux2009_2,Mandhour2020} can be attributed to the crossover of propagating to evanescent incident wavefunction.

The retention of the Klein tunneling behavior from $d=30$ [\cref{fig:2}(a)] to $d=100$, and subsequently to the \textit{n-p} limit is in \cref{fig:3} by setting $t=A_B$, $A_R=B_R=0$ in \cref{eqn:wavefun} s.t.
\begin{align}
    T_{\textit{np}}&=\frac{4\lambda' q_yk_y\varepsilon_\mbf{q}\varepsilon_\mbf{k}}{\lambda\lr{\abs{\lambda k_y\varepsilon_\mbf{q}+\lambda'\varepsilon_\mbf{k}q_y}^2+U^2_0(k^2_x+\Delta)^2}},
    \label{eqn:T_NP}
\end{align}
with the conservation of pseudospin dictating $\abs{r}^2 + \abs{t}^2 \lambda' q_{y}\varepsilon_\mbf{k}/(\lambda k_{y}\varepsilon_\mbf{q}) = 1$ and by neglecting the evanescent wave contributions.

\begin{figure} [t]
    \centering
    \includegraphics[width=0.45\textwidth]{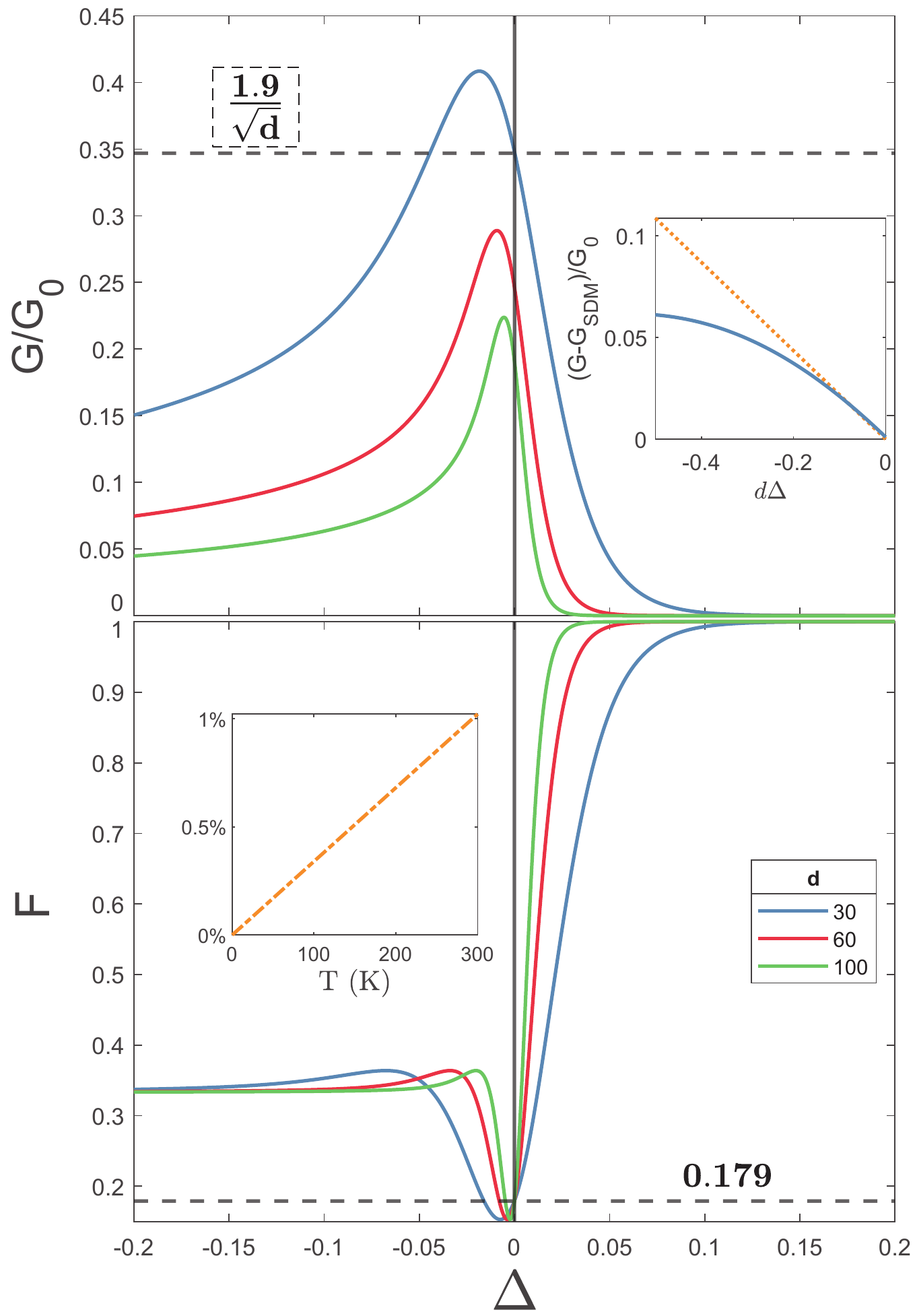}
    \caption{The conductance $G/G_0$ (top) and Fano factor $F$ (bottom) [\cref{eqn:G_F}] are plotted respectively across $\Delta$ with $d=30$ (blue), $d=60$ (red) and $d=100$ (green). 
    (Top): The broken line is the SDM conductance value, i.e. $G_{SDM}/G_0=1.9/\sqrt{d}$.
    The top inset shows the relationship between $(G-G_{\text{SDM}})/G_0$ (blue solid line) with $d=30$ against its linear approximation (orange dotted line) with $-0.5 \leq d\Delta \leq 0$ in the band inversion phase.
    (Bottom): The broken line is the SDM Fano factor, i.e. $F\approx 0.179$.
    The bottom inset shows the thermal noise correction up to $300K$ for all $\Delta$.
    } 
    \label{fig:4}
\end{figure}

We now investigate the quantum transport signatures, i.e., conductance $G$ and Fano factor $F$ [\cref{fig:4}]. 
In the zero-temperature limit, $G$ and $F$ can be expressed as
\begin{align}
    G= G_0 \int T \text{d}k_x;\quad F = \frac{S_I}{G},\label{eqn:G_F}
\end{align}
with $G_0 = ge^2/(2\pi h)$, $g$ is the degeneracy term (spin and valley), $e$ is the electronic charge, $h$ is the Planck's constant, $S_I=\int T(1-T) \text{d}k_x$ is the current noise and $T$ is obtained from \cref{eqn:Trans_NPN} in the grazing energy ($\varepsilon_\mbf{k}/U_0 \rightarrow \gg 1$) and tall barrier ($L_x \gg d$) limit \cite{Tworzydo2006} s.t.
\begin{align}
    T \rightarrow \sech^2((k^2_x +\Delta) d),
    \label{eqn:T_tall}
\end{align}
with $\varepsilon_\mbf{k} \rightarrow U_0$, $\varepsilon_\mbf{q} \rightarrow 0$, $q^2_y \rightarrow - (k^2_x +\Delta)^2$, and $k^2_y \rightarrow U^2_0-(k^2_x +\Delta)^2$.

The conductance $G/G_0$ and the Fano factor $F$ at $d=30$ (blue), $d=60$ (red) and $d=100$ (green) are shown in \cref{fig:4}.
Away from the semi-Dirac point with ($G/G_0 \approx 1.9/\sqrt{d}$, $F\approx0.179$) \cite{Zhai2014}, the values converge towards the insulating ($\Delta>0$) with ($G/G_0 = 0$, $F\rightarrow1$) and band inversion ($\Delta<0$) phase with ($G/G_0=1/d$ [see inset of top figure], $F\rightarrow1/3$) \cite{Tworzydo2006} respectively.
Strikingly, the convergence of the band inversion phase to the sub-Poissonian value in graphene \cite{Tworzydo2006} and disordered metals \cite{Nagaev1992,Beenakker1992} can be understood by observing that \cref{eqn:T_tall} captures only the carriers along the $\varepsilon_\mbf{k}/U_0 \approx1$ line [see \cref{fig:2}(a)].
However, this breaks down in the \textit{n-p} limit with vanishing $T$.

\begin{table*}[t]
\caption{Calculated $\Delta$ and Fano factors $F$ using material parameters from semi-Dirac materials with $\alpha \sim 1/m^*$, $\beta \sim v_y$, $k_0\sim m^*v_x$ and $\varepsilon_0 \sim m^*v^2_x$ in \cref{eqn:H_full}.
Both the $\Delta/\varepsilon_0$ and $F$ values are taken from the insulating and band inversion phase and are calculated using $d=30$ s.t. the width is approximately $2$ to $10$ nm. 
\label{tab:table1}}
\begin{ruledtabular}
\begin{tabular}{ccccccccc}
\textrm{Material}
&\textrm{Method}
&$m^*$ ($m_e$)
&($v_x$,$v_y$) ($10^6$m$/$s)
&$k_0$ ($\text{\r{A}}^{-1}$)
&$\varepsilon_0$ (eV)
&$\Delta/\varepsilon_0$
&$F$\\
\colrule
Black \ch{P} \cite{Baik2015,Kim2017} & Doping
&$1.42$
&($0.86,0.28$-$0.56$\footnote{Values from both the insulating and band inversion phase\cite{Baik2015}})
&$1.37$
&$5.065$
&($0.036$, -$0.006$)
&($0.733$, $0.154$)\\
$5$L black \ch{P} \cite{Ghosh2016,Baik2015} & E-Field
&$1.2$
& ($0.202$,$0.253$)
&$0.436$
&$0.580$
&($0.483$, -$0.362$\footnote{Linearly extrapolated from insulating phase \cite{Ghosh2016}})
&($1$, $0.334$)
\\
Bilayer \ch{P} \cite{Zhang2015,WangJAP2015}
&Strain
&$1.25$
&($0.562$, $0.75$\footnote{Approximated from single layer fermi velocities \cite{Yuan2015}})
&$0.607$
&$1.545$
&($0.044$, -$0.03$)
&($0.822$, $0.27$)\\
\end{tabular}
\end{ruledtabular}
\end{table*}

$G$ in \cref{eqn:G_F} can be conveniently utilized as a litmus test of the underlying band structure by linearly expanding \cref{eqn:T_tall} around the semi-Dirac phase ($\Delta=0$) with $T \approx T(\Delta=0) - \tilde{T}$ where $T(\Delta=0)$ is the tunneling probability at the semi-Dirac point, and 
\begin{align}
 \tilde{T} = 2d\Delta \tanh(k^2_x d)\sech^2(k^2_x d).
\end{align}
Correspondingly, the deviation of $G$ away from around the semi-Dirac point with universal conductance $G_{\text{SDP}}/G_0=1.9/\sqrt{d}$ is:
\begin{align}
    \frac{G-G_\text{SDP}}{G_0} \approx  - 2d\Delta \int\left(\tanh(k^2_x d)\sech^2(k^2_x d)\right)dk_x. \label{eqn:local_approx}
\end{align}
The range of validity of \cref{eqn:local_approx} [orange dotted line] against the actual deviation, i.e. $(G-G_\text{SDP})/G_0$ [blue solid line] at $d=30$ is shown in the top figure inset of \cref{fig:4}.

Elevated temperature can affect noise measurement by broadening the current fluctuations through thermal excitation, which effectively increase the current noise, $S_I$ and $F$ by approximately $1\%$ at $\mathrm{T}=300K$ [inset of \cref{fig:4} in bottom figure].
This is accounted by the measured averaged Fano factor $\mathcal{F}$, which deviates from the true $F$ with the inclusion of the $\Delta$-invariant thermal noise \cite{Khlus1987} s.t.
\begin{align}
    \mathcal{F} \approx \lr{\coth\lr{\frac{U_0}{2k_B\mathrm{T}}} - \frac{2k_B\mathrm{T}}{U_0}}F,\label{eqn:Thermal_noise}
\end{align}
where $U_0$ plays the role of $eV$ here, with $V$ being the applied voltage.
This approximation reduces to $S_T = 4k_B\mathrm{T} G = 0.0082 < 1\%$ if $T$ is highly dependent on $\varepsilon_\mbf{k}$ [see \cref{eqn:Trans_NPN}] and $eV=U_0 \ll k_B\mathrm{T}$.
Hence, the miniature contributions of the thermal noise in systems govern by \cref{eqn:H_full} removes the need for filtering thermal noise in electronics noise measurements as $\mathcal{F} \approx F$.

The $F$ values for SDMs in \cref{tab:table1} are calculated using \cref{eqn:G_F} with the extracted material coefficients.
Finally, we briefly comment on the experimental relevance of the model parameter, $\Delta$, which plays the important role in determining the band topology of 2D SDMs. The $\Delta$ term can be externally tuned by a large variety of methods, such as varying the dopant distance \cite{Baik2015}, doping density \cite{Kim2017}, applied electric field \cite{Baik2015,Ghosh2016} or strain \cite{Montambaux2009,Zhong2017,WangJAP2015,Zhang2015}.
However, cautionary measures should be taken to preserve the space-time inversion symmetry \cite{Ahn2019} to observe the phase transitions.
Counterexamples are observed either by straining perpendicular to the valleys \cite{WangJAP2015} or lacking the required symmetry such as in monolayer arsenene or non Tb-stacked phosphorous \cite{Wang2016,Zhang2017}.

In conclusion, we study the quantum transport and shot noise signature of a $2$D semi-Dirac system along the relativistic dispersion direction.
The quantum tunneling spectrum exhibits a peculiar coexistence of massless and massive Dirac quasiparticles with full and oscillating transmission probabilities, respectively. 
$2$D semi-Dirac system thus offer a versatile material platform to study the Klein tunneling phenomena.

We further obtain the ballistic tunneling conductance $G$ and the Fano factors $F$ values across different $\Delta$, namely ($G/G_0 \rightarrow 1/d$, $F\approx 1/3$), ($G/G_0 \rightarrow 1.9/\sqrt{d}$, $F\approx 0.179$ and ($G/G_0 \rightarrow 0$, $F\approx 1$) for the band inversion, semi-Dirac, and the band insulating phases, respectively.
The conductance and shot noise shall provide useful signatures to experimentally probe the phase transitions of a 2D semi-Dirac system.


W.J.C acknowledge MOE PhD RSS. Y.S.A is supported by the Singapore Ministry of Education Academic Research Fund Tier $2$ (MOE-T$2$EP$50221$-$0019$). L.K.A is supported by A*STAR AME IRG (A$2083$c$0057$). 

\section*{Author Declaration}
\subsection*{Conflict of Interest}
The authors have no conflicts to disclose.
\subsection*{Author Contributions}
\textbf{Wei Jie Chan}: Investigation (lead); Methodology (equal); Writing original draft (lead). 
\textbf{Lay Kee Ang}: Supervision (equal); Writing-review \& editing (equal).
\textbf{Yee Sin Ang}: Conceptualization (lead); Methodology (equal); Supervision (equal); Writing-review \& editing (equal).
\subsection*{Data Availability}
The data that support the findings of this study are available from the corresponding author upon reasonable request.

\bibliography{SDMKytunnel.bib}

\end{document}